\documentclass[draftclsnofoot,onecolumn,12pt]{IEEEtran}

\ifCLASSINFOpdf
\else
\fi

\usepackage{setspace}
\usepackage{bbm}
\usepackage[cmex10]{amsmath}
\usepackage{amssymb}
\usepackage{cite}
\usepackage{graphicx}
\usepackage{array,color}
\usepackage{amsmath}
\allowdisplaybreaks
\usepackage{stfloats}
\usepackage{graphicx}
\usepackage{epstopdf}
\usepackage{subfigure}
\usepackage{tabularx}
\usepackage{epsfig,epsf,color,balance,cite}
\usepackage{algorithmic}
\usepackage{algorithm}
\usepackage{url}
\usepackage{bm}
\usepackage{multirow}

\usepackage{amsthm}

\ifodd 0
\usepackage{soul}
\usepackage{color}
\setstcolor{red}

\newcommand{\rev}[1]{{\color{red}#1}} 
\newcommand{\del}[1]{\st{#1}} 

\newcommand{\com}[1]{\textbf{\color{red} (COMMENT: #1)}} 
\newcommand{\response}[1]{\textbf{\color{green} (RESPONSE: #1)}} 
\else

\newcommand{\rev}[1]{#1}

\newcommand{\del}[1]{}

\newcommand{\com}[1]{}
\newcommand{\comg}[1]{}
\newcommand{\response}[1]{}
\fi

\vspace{-0.4cm}
\title{\huge {Simultaneous Transmit Diversity and Passive Beamforming with Large-Scale Intelligent Reflecting Surface: Far-Field or Near-Field?}}

\author{ 
	Beixiong Zheng,~\IEEEmembership{Member,~IEEE} and Rui Zhang,~\IEEEmembership{Fellow,~IEEE} 
	\vspace{-1.75cm}
	
	\thanks{
		
		B. Zheng is with the School of Microelectronics, South China University of Technology, Guangzhou 511442, China (e-mail: bxzheng@scut.edu.cn). He was with the Department of Electrical and Computer Engineering, National University of Singapore, Singapore 117583.
		
		R. Zhang is with the Department of Electrical and Computer Engineering, National University of Singapore, Singapore 117583 (e-mail: elezhang@nus.edu.sg).
	}
}

\begin{document}
\maketitle
\begin{abstract}
Intelligent reflecting surface (IRS) has emerged as a cost-effective solution to enhance wireless communication performance via passive signal reflection. Existing works on IRS have mainly focused on investigating IRS's passive beamforming/reflection design to boost the communication rate for users assuming that their channel state information (CSI) is fully or partially known. However, how to exploit IRS to improve the wireless transmission reliability without any CSI, which is typical in high-mobility/delay-sensitive communication scenarios, remains largely open.   
In this paper, we study a new IRS-aided communication system with the IRS integrated to its aided access point (AP) to achieve both functions of transmit diversity and passive beamforming simultaneously. Specifically, we first show an interesting result that the IRS's passive beamforming gain in any channel direction is invariant to the common phase-shift applied to all of its reflecting elements.
Accordingly, we design the common phase-shift of IRS elements to achieve transmit diversity at the AP side without the need of any CSI of the users. In addition, we propose a practical method for the users to estimate the CSI at the receiver side for information decoding.  Meanwhile, we show that the conventional passive beamforming gain of IRS can be retained for the other users with their CSI known at the AP.
Furthermore, we derive the asymptotic performance of both IRS-aided transmit diversity and passive beamforming in closed-form, by considering the large-scale IRS with an infinite number of elements.
Numerical results validate our analysis and show the performance gains of
the proposed IRS-aided simultaneous transmit diversity and passive beamforming scheme over other benchmark schemes.
\end{abstract}
\vspace{-0.35cm}
\begin{IEEEkeywords}
	Intelligent reflecting surface (IRS), large-scale IRS, near-field modeling,
	space-time code, transmit diversity, passive beamforming, channel estimation, performance analysis.
\end{IEEEkeywords}
\IEEEpeerreviewmaketitle

\section{Introduction}
Leveraging the recent advance in digitally-controlled metasurfaces,
intelligent reflecting surface (IRS) has emerged as an innovative technology to \rev{materialize the marvelous concept of ``smart and reconfigurable radio environment" via passive signal reflection \cite{wu2020intelligent,zheng2021survey,qingqing2019towards,Renzo2019Smart,Huang2020Holographic,basar2019wireless}}, which has drawn increasing attention from both
academia and industry.
Specifically, IRS is composed of a large number of passive reflecting elements, each of which can be independently tuned to change the amplitude/phase-shift of the incident signal in real time \rev{with ultra-low power consumption \cite{Huang2019Reconfigurable,Wu2019TWC}.}
By smartly tuning its massive reflecting elements, IRS is able to proactively reshape the wireless propagation channel in favor of signal transmission, in contrast to only adapting to the random and time-varying wireless channel by traditional transceiver techniques.
As such, IRS opens up various new and appealing functions for wireless communications (such as passive beamforming, interference suppression, channel statistics refinement, etc. \cite{wu2020intelligent}) to effectively combat against the wireless channel impairments and thus
enhance the communication performance drastically.
Furthermore, since IRS only passively reflects the ambient radio signal and is free of radio frequency (RF) chains, it requires only low hardware cost and power consumption. Besides, its practical features such as low profile, lightweight,
and conformal geometry also facilitate the flexible and large-scale deployment of IRS in future wireless networks \cite{wu2020intelligent,zheng2021survey,qingqing2019towards}.


The above appealing features/functions of IRS have spurred growing interest in studying its performance gains under different wireless
system setups, such as orthogonal frequency division multiplexing (OFDM) \cite{zheng2019intelligent,zheng2020intelligent,yang2019intelligent,Zheng2020Fast}, \rev{multi-antenna communication \cite{wei2021channel,huang2020reconfigurable,zhang2019capacity}, multi-cell network \cite{Pan2020Multicell,Xie2021Max,Luo2021Reconfigurable,lyu2021hybrid},  double-/multi-IRS network \cite{Zheng2020DoubleIRS,zheng2020efficient,mei2021intelligent,huang2021Multi-Hop,zheng2022intelligent},} relaying communication \cite{zheng2021irs,Yildirim2021Hybrid,nguyen2021hybrid}, multiple access \cite{Zheng2020IRSNOMA,Guo2021Intelligent,Zuo2021Reconfigurable},
among others.
However, most of the existing works on IRS have focused on its passive beamforming/reflection design for enhancing the communication performance for users with their channel state information (CSI) perfectly or partially known (see,
e.g., \cite{wu2020intelligent,zheng2021survey} and the references therein).
This thus requires efficient channel estimation and/or beam training to reduce the channel training/feedback overhead. 
However, due to the practically very large number of passive elements at each IRS that do not possess transmitting/receiving capabilities, the widely-adopted ``all-at-once" IRS channel/beam training scheme (see,
e.g., \cite{zheng2019intelligent,zheng2020intelligent,yang2019intelligent,Zheng2020Fast,You2020Fast,wei2021channel}) may incur prohibitive overhead that is generally proportional to the number of reflecting elements, thus causing a long delay prior to data transmission. Although this issue may be mild for quasi-static/low-mobility users for which their CSI can be acquired within sufficiently long channel coherence interval, \rev{it becomes more severe for delay-sensitive/short-packet transmissions, which are typical for scenarios with high-mobility users such as vehicular communications \cite{Chen2022Robust,Sun2021channel,huang2021transforming,Al-Hilo2022Reconfigurable}.  
In such scenarios, beam misalignment is more likely to happen for IRS passive beamforming, making the beam tracking problem even more challenging to tackle.}
In view of the above issues and limitations of IRS passive beamforming, it is imperative to design new and efficient schemes for IRS to enhance the wireless transmission reliability for high-mobility and delay-sensitive communications with no or little CSI.

%

\begin{figure}[!t]
	\centering
	\includegraphics[width=3.3in]{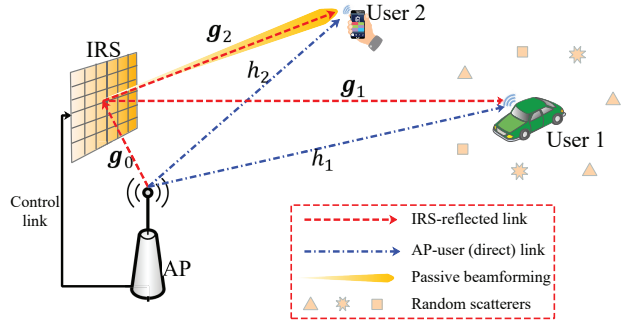}
	\setlength{\abovecaptionskip}{-3pt}
	\caption{An IRS-integrated AP in downlink communication with both high-mobility (User 1) and low-mobility (User 2) users.}
	\label{system}
	\vspace{-0.7cm}
\end{figure}
Motivated by the above, we consider in this paper a new IRS-aided downlink communication system with the IRS integrated to its aided access point (AP) to achieve both transmission reliability and passive beamforming for high-/low-mobility users at the same time without and with CSI, respectively, as shown in Fig.~\ref{system}.
Under this system setup, we unlock a new function of IRS to help achieve transmit diversity at the AP in addition to the conventional function of passive beamforming.
Specifically, we design the IRS's common phase-shift that is applied to all of its reflecting elements to achieve transmit diversity at the AP side, without the need of any CSI of the high-mobility users. As such, this new function of IRS is practically appealing for delay-sensitive and/or short-packet transmissions (e.g., the transmission of control information in high-mobility/fast-fading communication scenarios) to achieve a transmit diversity gain and thereby improve the transmission reliability.
Meanwhile, the conventional passive beamforming function of IRS can also be achieved simultaneously to boost the communication rate of low-mobility users with known CSI at the AP as they are more tolerable to the training/feedback delay, by leveraging an interesting fact that the IRS passive beamforming gain in any given channel is invariant to the common phase-shift applied to its reflecting elements for achieving the transmit diversity. 

Moreover, we consider a new IRS-aided system with the IRS integrated to its aided AP, for the ease of implementing the real-time control of IRS reflection as well as minimizing its distance with the AP to reduce the propagation loss in its reflected link with the users. However, due to the co-located IRS and AP, the size of IRS can no longer be ignored in its channel modeling and the simple far-field propagation model assumed in most existing works on IRS becomes inaccurate. In this case, the IRS channel modeling needs to account for the variations in the incident signal's both strength and angle-of-arrival (AoA) across different reflecting elements at the IRS \cite{feng2021wireless}.
The main contributions of this paper are summarized as follows.
\begin{itemize}
	\item First, inspired by the celebrated Alamouti's scheme \cite{alamouti1998simple}, we propose a new space-time code design  for the joint AP and IRS transmission/reflection to achieve transmit diversity without the need of any CSI. Specifically, with one ``active" antenna equipped at the AP, the ``passive" IRS dynamically tunes its common phase-shift based on the phase difference between two consecutive modulated symbols that are transmitted by the AP. In addition,
	 a practical method is proposed for the users to estimate the CSI at the receiver side for information decoding.
	Then, based on the element-wise IRS channel modeling, we analyze the transmit diversity performance of the proposed design and derive the results in closed-form that provide further insights. In particular, our asymptotic result reveals that an infinitely large IRS can reflect at most half of the power transmitted by an isotropic source (the AP) to help achieve a transmit diversity gain of order two.
	\item Next, we show that with the proposed transmit diversity scheme by tuning the IRS common phase-shift, the IRS passive beamforming gain can be simultaneously achieved for users with their CSI known at the AP. To this end, we derive the tight lower- and upper-bounds for the maximum passive beamforming gain in closed-form. Moreover, we analyze the asymptotic performance and show that the IRS passive beamforming gain increases with the number of reflecting elements and eventually converges to a constant
	value. \rev{This result is in sharp contrast to the so-called ``squared power gain'' of IRS passive beamforming under the far-field assumption, which increases with the number of IRS reflecting elements {\it quadratically} without any upper bound \cite{Wu2019TWC}.} In addition, we show that with an infinitely large IRS,
	the asymptotic passive beamforming gain decays with the propagation distance only, instead of its square as in the free-space path loss model.
	\item Finally, we provide extensive numerical results to validate the performance of the proposed scheme of simultaneous transmit diversity and passive beamforming for the IRS-aided AP.
	It is shown that the proposed scheme achieves better performance than other benchmark schemes.
	Besides, we demonstrate the necessity of proper IRS channel modeling for accurately characterizing both the transmit diversity and passive beamforming performance in
	the considered IRS-aided downlink communication.
\end{itemize}

The rest of this paper is organized as follows. Section~\ref{sys} presents the system model with the integrated AP
and IRS, and introduces the new scheme of simultaneous transmit diversity and passive beamforming for the IRS-aided downlink communication. In Section~\ref{diversity}, we propose the new IRS-aided transmit diversity scheme and analyze its performance. 
In Section~\ref{Beamforming}, we design the IRS passive beamforming with the given transmit diversity scheme and analyze its performance. 
Simulation results are presented in Section \ref{Sim} to evaluate the performance of the
proposed designs and validate our analytical results. Finally, conclusions are drawn in Section~\ref{conlusion}.

\emph{Notation:} 
Upper-case and lower-case boldface letters denote matrices and column vectors, respectively.
Upper-case calligraphic letters (e.g., $\cal{M}$) denote discrete and finite sets.
Superscripts ${\left(\cdot\right)}^{T}$, ${\left(\cdot\right)}^{H}$, ${\left(\cdot\right)}^{*}$, ${\left(\cdot\right)}^{-1}$, and ${\left(\cdot\right)}^{\dagger}$ stand for the transpose, Hermitian transpose, conjugate, matrix inversion, and pseudo-inverse operations, respectively.
${\mathbb C}^{a\times b}$ denotes the space of ${a\times b}$ complex-valued matrices.
For a complex-valued vector $\bm{x}$, $\lVert\bm{x}\rVert_{\ell}$ denotes its $\ell $-norm,
$\angle (\bm{x} )$ returns the phase of each element in $\bm{x}$,
and ${\rm diag} (\bm{x})$ returns a diagonal matrix with the elements in $\bm{x}$ on its main diagonal.
$|\cdot|$ denotes the absolute value if applied to a complex-valued number or the cardinality if applied to a set,
and ${\mathbb E}\{\cdot\}$ stands for the statistical expectation.
${\bm I}$ and ${\bm 0}$ denote an identity matrix and an all-zero matrix, respectively, with appropriate dimensions.
The distribution of a circularly symmetric complex Gaussian (CSCG) random vector with zero-mean and covariance matrix ${\bm \Sigma}$ is denoted by ${\mathcal N_c }({\bm 0}, {\bm \Sigma} )$; and $\sim$ stands for ``distributed as".
\section{System Model and Proposed Scheme}\label{sys}

\subsection{System Model}
As shown in Fig.~\ref{system}, we consider an IRS-aided downlink communication system, where an auxiliary IRS is co-located/integrated with a single-antenna AP for enhancing its information transmission to the users\footnote{\rev{The results of this paper can be extended to the multi-antenna setup by jointly designing the multi-antenna AP's active beamforming and the IRS's passive beamforming \cite{Wu2019TWC} to serve multiple users at the same time over a given frequency band for the passive beamforming mode, while applying the more general space-time code design \cite{jafarkhani2005space} to the multi-antenna AP and the multi-IRS for the transmit diversity mode.}}.
Specifically, the IRS consisting of $N$ passive reflecting elements
 is connected via a reliable wired link to the AP for real-time control, thus referred to as the ``IRS-integrated AP" in this paper. Under this setup, the AP takes over the role of the conventional IRS controller for adjusting the IRS reflection in real time.

At the IRS-integrated AP, we consider two transmission modes referred to as the ``transmit diversity" and ``passive beamforming" modes, respectively, to support multiple single-antenna users, depending on their communication requirements and/or channel conditions. 
For the purpose of investigating the new scheme of simultaneous transmit diversity and passive beamforming, we focus on the simple case of two users only in this paper\footnote{\rev{The results of this paper can be extended to the  case of more than two users provided that the transmit diversity mode is applied over one orthogonal frequency band only at any time, for transmitting the same information to one or more users (i.e., multi-casting); meanwhile, the passive beamforming mode is applied over other orthogonal frequency bands for other  users in e.g., IRS-aided orthogonal frequency-division multiple access (OFDMA) systems \cite{Yang2020IRS}).}}, which are assigned with two orthogonal frequency sub-bands to communicate with the AP. 
The channels in each sub-band are assumed to be frequency-flat (i.e., the narrow-band channel model is assumed).
For convenience, we assume that user~1 (high-mobility) and user~2 (low-mobility) are served by the IRS-integrated AP via the transmit diversity mode and passive beamforming mode over the two sub-bands, respectively.
Accordingly, we let ${{\bm g}}_0\in {\mathbb{C}^{N\times 1 }}$,
${{\bm g}}_k\in {\mathbb{C}^{N\times 1 }}$, and ${h}_k\in {\mathbb{C}}$
denote the baseband equivalent channels for the AP$\rightarrow$IRS, IRS$\rightarrow$user $k$, and AP$\rightarrow$user $k$ links, respectively, with $k=\{1,2\}$.
Moreover, we let ${\bm \theta}\triangleq[{\theta}_{1},{\theta}_{2},\ldots,{\theta}_{N}]^T$ denote
the IRS reflection vector,
where the reflection amplitudes of all passive reflecting elements are set to one or the maximum
value, i.e., $|{\theta}_{n}|=1, \forall n=1,\ldots,N$, to maximize the signal reflection power as well as ease the hardware implementation \cite{Zheng2020DoubleIRS,zheng2020efficient,mei2021intelligent,Wu2019TWC}. 
\vspace{-0.3cm}
\subsection{Simultaneous Transmit Diversity and Passive Beamforming}
Under the unit-modulus constraint, any IRS reflection vector ${\bm \theta}$ can be expressed as
\begin{align}\label{decomposition}
{\bm \theta}=e^{j\varphi}{\bar{\bm \theta}}\Leftrightarrow  {\theta}_{n}=e^{j\varphi}{\bar{\theta}}_{n}, \qquad \forall n=1,\ldots,N
\end{align}
where ${\bar{\bm \theta}}\triangleq[{\bar{\theta}}_{1},{\bar{\theta}}_{2},\ldots,{\bar{\theta}}_{N}]^T$ denotes the (passive) beamfoming vector with $|{\bar{\theta}}_{n}|=1, \forall n=1,\ldots,N$,
and $\varphi$ denotes the IRS's common phase-shift of all its reflecting elements.
As will be shown later in Section~\ref{Beamforming}, the IRS passive beamforming gain in any direction depends on ${\bar{\bm \theta}}$ only and is invariant to $\varphi$.
As such, based on \eqref{decomposition}, we propose to design the IRS common phase-shift $\varphi$ and the beamforming vector ${\bar{\bm \theta}}$ to achieve the transmit diversity and passive beamforming for user~1/user~2 without and with CSI, respectively. 

\begin{figure}[!t]
	\centering
	\includegraphics[width=4.5in]{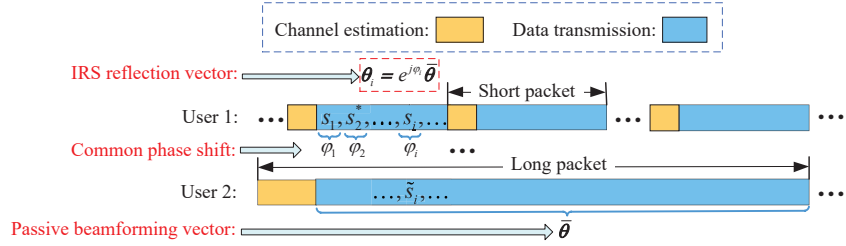}
	\setlength{\abovecaptionskip}{-3pt}
	\caption{Transmission protocol for IRS-aided simultaneous transmit diversity and passive beamforming.}
	\label{protocol}
	\vspace{-0.7cm}
\end{figure}
As shown in Fig.~\ref{protocol}, the proposed transmission protocol serves the two users simultaneously with short-packet
and long-packet transmissions, respectively. 
Specifically, for the former case with short packets, each transmission packet for user~1 consists of one short pilot sequence and subsequently one short data frame.
Moreover, we assume that within each short packet, the channels $\left\{{{\bm g}}_1, {h_1}\right\}$ of user~1 remain constant.
During each short-packet transmission,
the IRS common phase-shift $\varphi$ can be dynamically tuned on a per-symbol basis to help achieve the transmit diversity gain for user~1.
On the other hand, for the latter case with long packets,
each transmission packet for user~2 constitutes one long pilot sequence followed by one long data frame, as illustrated in Fig.~\ref{protocol}. 
Note that the channels $\left\{{{\bm g}}_2, {h_2}\right\}$ of user~2 are assumed to remain constant during each long packet. As such, the beamforming vector ${\bar{\bm \theta}}$ is fixed over each long packet to achieve a constant passive beamforming gain for user~2.

\vspace{-0.3cm}
\section{Transmit Diversity}\label{diversity}
In this section, we focus on the transmit diversity mode for user~1 to draw essential and useful insights. Specifically, we study the communication design and asymptotic performance for the IRS-aided transmit diversity scheme. 
\vspace{-0.3cm}
\subsection{Space-time Code Design}
Under the transmission protocol shown in Fig.~\ref{protocol} and inspired by the celebrated Alamouti's scheme \cite{alamouti1998simple}, we propose the IRS-aided transmit diversity scheme at the AP
 to serve user~1.
In particular, we consider that there is no CSI of user~1 available at the IRS-integrated AP; while user~1 can acquire the CSI at the receiver side for information decoding via a simple channel estimation method (to be specified in Section~\ref{CE}).
As such, the proposed transmit diversity mode at the IRS-integrated AP
 applies to both time-division duplexing (TDD) and frequency-division duplexing (FDD) systems without the need of assuming channel reciprocity and CSI feedback, which is thus highly appealing to the practical implementation in high-mobility/delay-sensitive communication scenarios, such as vehicle-to-everything (V2X) communication.
\rev{Note that different from the classic Alamouti's scheme \cite{alamouti1998simple} that requires two ``active" transmit antennas at the AP (referred to as AP antenna 1 and AP antenna 2 in the sequel, respectively) to achieve transmit diversity, our proposed IRS-integrated AP only consists of one ``active" transmit antenna (labeled as AP antenna 1) and one ``passive" IRS to achieve the transmit diversity function}, as shown in Fig.~\ref{system}.

\begin{table}[]
	\centering
	\caption{The Space-time Code Design for IRS-aided Transmit Diversity Scheme Versus Alamouti's Scheme.}\label{coding}
	\vspace{-0.5cm}
	\resizebox{0.8\textwidth}{!}{
	\begin{tabular}{c|cc||cc|}
		\cline{2-5}
		& \multicolumn{2}{c||}{IRS-aided transmit diversity scheme}                     & \multicolumn{2}{c|}{Alamouti's scheme \cite{alamouti1998simple}}           \\ 
		& \multicolumn{2}{c||}{(for the IRS-integrated AP)}                     & \multicolumn{2}{c|}{(for the two-antenna AP)}           \\ \cline{1-5} 
		\multicolumn{1}{|c|}{Symbol index} & \multicolumn{1}{c|}{AP antenna 1} & IRS common phase-shift & \multicolumn{1}{c|}{AP antenna 1} & AP antenna 2 \\ \hline
		\multicolumn{1}{|c|}{1} & \multicolumn{1}{c|}{$s_1$}    & $\varphi_1=\angle s_2-\angle s_1$~~~~~             & \multicolumn{1}{c|}{$s_1$}    & $s_2$    \\ \hline
		\multicolumn{1}{|c|}{2} & \multicolumn{1}{c|}{$-s_2^*$} & $\varphi_2=\angle s_2-\angle s_1+\pi$ & \multicolumn{1}{c|}{$-s_2^*$} & $s_1^*$ \\ \hline
	\end{tabular}
}
\vspace{-0.8cm}
\end{table}

During each short-packet transmission, the AP and IRS jointly encode consecutive pairs of modulated data symbols, which, in any give pair, are denoted by $s_1$ and $s_2$ according to a new space-time code design for user~1. For simplicity, we assume  $s_1$ and $s_2$ are independently drawn from an $M$-ary phase-shift keying (PSK) constellation.\footnote{\rev{The PSK modulation considered in this paper can be extended to the amplitude/phase-shift keying (APSK) modulation with additional information embedded in the (common) signal amplitude of $s_1$ and $s_2$.}}
The space-time code design for the joint AP and IRS transmission/reflection is shown in Table \ref{coding}.
Specifically, during the first symbol period, the AP transmits $s_1$ and the IRS sets its reflection vector as
${\bm \theta}_1=e^{j\varphi_1}{\bar{\bm \theta}}$ with the common phase-shift being the phase difference between the two modulated symbols, i.e., $\varphi_1=\angle s_2-\angle s_1$.
During the second symbol period, the AP transmits $-s_2^*$ and the IRS sets its reflection vector as 
${\bm \theta}_2=e^{j\varphi_2}{\bar{\bm \theta}}$ with the common phase-shift being $\varphi_2=\angle s_2-\angle s_1+\pi$. 
In Table~\ref{coding}, we show the space-time code design of the proposed transmit diversity scheme for the IRS-integrated AP versus the classic Alamouti's scheme for the two-antenna AP. It is observed that the same sequence $\left[s_1, -s_2^*\right]^T$ applies to AP antenna~1 in both schemes; however, in contrast to the sequence $\left[s_2, s_1^*\right]^T$ applied at AP antenna 2 in the Alamouti's scheme, the IRS tunes its common phase-shifts $\left\{\varphi_1, \varphi_2\right\}$ based on the phase difference between the two modulated symbols in the IRS-aided transmit diversity scheme.

According to the above, the received signal at user~1 during the first symbol period is written as
\begin{align}\label{S1}
y_1&=\sqrt{P_1}\left({h_1} +{{\bm g}}_0^T  {\rm diag}\left({\bm \theta}_1\right) {{\bm g}}_1\right)s_1+n_1\notag\\
&=\sqrt{P_1}\left({h_1} s_1 +{{\bm g}}_0^T  {\rm diag}\left({\bar{\bm \theta}}\right) {{\bm g}}_1 e^{j\varphi_1} s_1\right)+n_1\stackrel{(a1)}{=}\sqrt{P_1}\left({h_1} s_1 +{\bar g}_1 s_2\right)+n_1
\end{align}
where $P_1$ denotes the AP transmit power for user~1, ${\bar g}_1\triangleq {{\bm g}}_0^T  {\rm diag}\left({\bar{\bm \theta}}\right) {{\bm g}}_1$
denotes the effective complex-valued gain of the IRS-reflected (i.e., AP$\rightarrow$IRS$\rightarrow$user~1) channel,
$n_1$ is the zero-mean additive white Gaussian noise (AWGN) with variance $\sigma^2$ at user~1, and $(a1)$ holds since
$s_2=s_1 e^{j (\angle s_2-\angle s_1)}=e^{j\varphi_1} s_1$. On the other hand, during the second symbol period, the received signal at user~1 is given by
\vspace{-0.2cm}
\begin{align}\label{S2}
y_2&=-\sqrt{P_1}\left({h_1} +{{\bm g}}_0^T  {\rm diag}\left({\bm \theta}_2\right) {{\bm g}}_1\right)s_2^* +n_2\notag\\
&=\sqrt{P_1}\left(-{h_1}s_2^* -{{\bm g}}_0^T  {\rm diag}\left({\bar{\bm \theta}}\right) {{\bm g}}_1 e^{j\varphi_2} s_2^*\right) +n_2\stackrel{(a2)}{=}\sqrt{P_1}\left(-{h_1} s_2^* +{\bar g}_1 s_1^*\right)+n_2
\end{align}
where $n_2$ is the zero-mean AWGN with variance $\sigma^2$ at user~1 and $(a2)$ holds since $s_1^*=-s_2^*e^{j (\angle s_2-\angle s_1+\pi)}=-e^{j\varphi_2}s_2^*$.
\vspace{-0.3cm}
\subsection{Decoding Design}
Since we are interested in decoding $s_1$ and $s_2$ at user~1, we let ${\bm y}\triangleq\left[y_1, y_2^*\right]^T$ denote the received signal vector for each transmitted pair of $s_1$ and $s_2$. Accordingly, \eqref{S1} and \eqref{S2} can be rewritten in a compact form as
\vspace{-0.2cm}
\begin{align}\label{S1S2}
{\bm y}=\sqrt{P_1}\underbrace{\begin{bmatrix}
	{h_1}  &{\bar g}_1\\
	{\bar g}_1^*  &-{h_1}^*\end{bmatrix}}_{{\bm H}}
\underbrace{\begin{bmatrix}
	s_1 \\
	s_2 \end{bmatrix}}_{{\bm s}}+\underbrace{\begin{bmatrix}
	n_1 \\
	n_2 \end{bmatrix}}_{{\bm n}} 
\end{align}
where ${\bm H}$ denotes the equivalent channel matrix, ${\bm s}$ is the modulated symbol vector, and ${\bm n}$ is the AWGN vector with ${\bm n}\sim {\mathcal N_c }({\bm 0}, \sigma^2{\bm I})$.
It can be verified that the two columns of the square matrix ${\bm H}$ in \eqref{S1S2} are orthogonal to each other and the received signal model in \eqref{S1S2} bears the same form as that in the classic Alamouti's scheme \cite{alamouti1998simple}.
As such, we left-multiply the received signal vector ${\bm y}$ in \eqref{S1S2} by ${\bm H}^H$, which yields \cite{alamouti1998simple}
\begin{align}\label{decoupled}
{\bar{\bm y}}={\bm H}^H{\bm y}=\sqrt{P_1}{\bm H}^H {\bm H}{\bm s}+{\bar{\bm n}}
\end{align}
with ${\bar{\bm y}}\triangleq\left[{\bar y}_1,{\bar y}_2\right]^T$ and ${\bar{\bm n}}\triangleq{\bm H}^H {\bm n}$. Moreover, we obtain  
\begin{align}\label{Heq}
{\bm H}^H {\bm H}=\begin{bmatrix}
|{h_1}|^2+|{\bar g}_1|^2  &0\\
0  &|{h_1}|^2+|{\bar g}_1|^2\end{bmatrix}
\end{align}
and it can be verified that ${\bar{\bm n}}$ is the equivalent AWGN vector with ${\bar{\bm n}}\sim {\mathcal N_c }({\bm 0}, (|{h_1}|^2+|{\bar g}_1|^2)\sigma^2{\bm I})$.
Accordingly, the received signal-to-noise ratio (SNR) at user~1 is given by \cite{alamouti1998simple}
\begin{align}\label{SNR}
\gamma=\frac{P_1(|{h_1}|^2+|{\bar g}_1|^2)}{\sigma^2}.
\end{align}
\subsection{Channel Estimation}\label{CE}
According to \eqref{S1S2}-\eqref{Heq}, user~1 only needs to acquire the effective complex-valued gains of the direct and reflected channels, i.e., $\{{h_1}, {\bar g}_1\}$ for decoding the information of each pair of $s_1$ and $s_2$ transmitted by the IRS-integrated AP. \rev{In the following, we propose a simple yet efficient channel estimation method at user~1 tailored for the IRS-aided transmit diversity scheme.}

\rev{During the channel estimation stage, the IRS dynamically tunes its common phase-shift over different
pilot symbol periods to facilitate the channel estimation of $\{{h_1}, {\bar g}_1\}$} and the received pilot
signal at user~1 can be expressed as
\begin{align}\label{rec_pilot}
z_{l}&=\sqrt{P_1}\left({h_1} +{{\bm g}}_0^T  {\rm diag}\left({\bm \theta}_{l}\right) {{\bm g}}_1\right)x'_{l}+w_{l}\notag\\
&=\sqrt{P_1}\left({h_1} +{\bar g}_1 e^{j\varphi'_{l}} \right)x'_{l}+w_{l}, \qquad  l=1,\ldots, L
\end{align}
where ${\bm \theta}_{l}=e^{j\varphi'_{l}}{\bar{\bm \theta}}$ is the IRS reflection vector with $\varphi'_{l}$ being the training common phase-shift during pilot symbol $l$, $x'_{l}$ is 
the pilot symbol transmitted by the AP, $w_{l}$ is the zero-mean AWGN with variance $\sigma^2$,
and $L$ is the total number of plot symbols. \rev{Accordingly, by simply setting the same pilot symbol $x'_{l}=1, \forall l$ over different pilot symbol periods and 
stacking the $L$ received plot symbols at user~1 into ${\bm z}=\left[z_1,\ldots, z_L\right]^T$,} we obtain
\begin{align}\label{rec_pilot2}
{\bm z}=\sqrt{P_1}
\underbrace{\begin{bmatrix}
	1  &e^{j\varphi'_1}\\
	\vdots  &\vdots\\
	1  &e^{j\varphi'_{L}}\end{bmatrix}}_{\bm \Phi}
\begin{bmatrix}
{h_1}\\
{\bar g}_1
\end{bmatrix}
+{\bm w}
\end{align}
where ${\bm \Phi}$ denotes the IRS training reflection matrix and ${\bm w}=\left[w_1,\ldots, w_L\right]^T$
is the corresponding AWGN vector. By properly designing the training common phase-shifts $\left\{\varphi'_{l}\right\}_{l=1}^{L}$ such that ${\rm rank}( {\bm \Phi})=2$, the least-square (LS) estimates of ${h_1}$ and ${\bar g}_1$ based on \eqref{rec_pilot2} are given by
\begin{align}
\begin{bmatrix}
{\hat{h_1}}\\
{\hat{\bar g}_1}
\end{bmatrix}=\frac{1}{\sqrt{P_1}} {\bm \Phi}^{\dagger} {\bm z}= \begin{bmatrix}
{h_1}\\
{\bar g}_1
\end{bmatrix}+\frac{1}{\sqrt{P_1}} {\bm \Phi}^{\dagger}{\bm w}
\end{align}
where ${\bm \Phi}^{\dagger}=\left({\bm \Phi}^H{\bm \Phi}\right)^{-1} {\bm \Phi}^H$. \rev{Note that $L\ge 2$ is required for the pilot sequence to ensure the existence of ${\bm \Phi}^{\dagger}$ with ${\rm rank}( {\bm \Phi})=2$, which is independent of
the number of IRS reflecting elements $N$ and thus can be very small. For example, given $L=2$ as the minimum training overhead in the IRS-aided transmit diversity scheme, we can design the training reflection matrix as ${\bm \Phi}=\begin{bmatrix}
	1&-1\\
	1&1
	\end{bmatrix}$ with $\varphi'_1=0$ and $\varphi'_2=-\pi$.}
\subsection{Performance Analysis}\label{Analysis}
\begin{figure}[!t]
	\centering
	\includegraphics[width=2.8in]{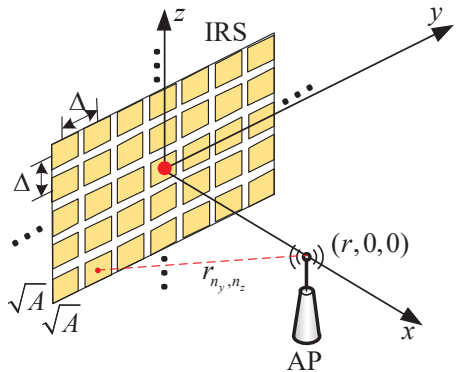}
	\setlength{\abovecaptionskip}{-3pt}
	\caption{Geometry relationship between the AP and IRS in the 3D Cartesian coordinate system.}
	\label{Geometry}
	\vspace{-0.7cm}
\end{figure}

In this subsection, we analyze the average channel gain and symbol error rate (SER) of the IRS-aided transmit diversity scheme. 
We assume that the IRS is equipped with a uniform planar array (UPA) that is placed on the $y$-$z$ plane and centered at the origin in the three-dimensional (3D) Cartesian coordinate system shown in Fig.~\ref{Geometry}. We let $N_y$ and $N_z$ denote the numbers
of reflecting elements along the $y$- and $z$-axis, respectively, and thus we have $N=N_y\times N_z$. 
We consider the equal spacing for the IRS elements along the $y$- and $z$-axis, which is denoted by $\Delta$. 
As illustrated in Fig.~\ref{Geometry},
the physical size of each reflecting element is denoted as $\sqrt{A}\times\sqrt{A}$ with $\sqrt{A}\le \Delta$, and we define $\xi\triangleq\frac{A}{\Delta^2}\le 1$ as the array occupation ratio of the effective IRS area to the overall UPA area.

For notational convenience, we assume that $N_y$ and $N_z$ are odd numbers. The central location of the $(n_y,n_z)$-th IRS element is denoted by ${\bm p}_{n_y,n_z}=[0,n_y\Delta,n_z\Delta]$, where $n_y=0,\pm1,\ldots, \pm(N_y-1)/2$ and $n_z=0,\pm1,\ldots, \pm(N_z-1)/2$. 
For ease of exposition, we assume that the AP antenna is placed along the $x$-axis and its location is denoted by ${\bm p}_{A}=[r,0,0]$ with $r>0$.
Accordingly, the distance between the AP and the center of the $(n_y,n_z)$-th IRS element is given by
\begin{align}
r_{n_y,n_z}=\left\|{\bm p}_{A}-{\bm p}_{n_y,n_z}\right\|_2=\sqrt{r^2+ n_y^2\Delta^2+n_z^2\Delta^2}=r\sqrt{1+ (n_y^2+n_z^2)\epsilon^2}
\end{align}
where we have $r= r_{0,0}$ being the shortest distance between the AP antenna and the IRS, and $\epsilon\triangleq\frac{\Delta}{r}$ with $\epsilon\ll 1$ in practice, which is due to the fact that the element spacing is typical on the  order of sub-wavelength at the signal carrier frequency.

\rev{Due to the short distance between the AP and IRS, the AP$\rightarrow$IRS channel ${{\bm g}}_0$ is modeled as the free-space line-of-sight (LoS) channel in this paper.}
Furthermore, \rev{by taking into account the variations in both path loss and projected
aperture across different reflecting elements}, the channel power
gain between the (isotropic) AP antenna and the  $(n_y,n_z)$-th IRS element can be characterized as \cite{lu2021communicating,lu2021multi,feng2021wireless}
\begin{align}\label{AP_IRS}
a_{n_y,n_z}\hspace{-0.1cm}=\underbrace{\frac{1}{4\pi \left\|{\bm p}_{A}-{\bm p}_{n_y,n_z}\right\|_2^2}}_{\rm Free-space~path~loss}  \underbrace{A\frac{({\bm p}_{A}-{\bm p}_{n_y,n_z})^T{\bm e}_x}{\left\|{\bm p}_{A}-{\bm p}_{n_y,n_z}\right\|_2}}_{\rm Projected~aperture}=\frac{Ar}{4\pi r_{n_y,n_z}^3}=\frac{A}{4\pi r^2\left(1+ (n_y^2+n_z^2)\epsilon^2\right)^{3/2}}
\end{align}
where ${\bm e}_x=[1,0,0]$ denotes the unit vector along the $x$-axis, which is also the normal vector of each IRS element placed on the $y$-$z$ plane. Based on the channel gain model in \eqref{AP_IRS}, the channel coefficient between the (isotropic) AP antenna and the  $(n_y,n_z)$-th IRS element in ${{\bm g}}_0$ is given by
\begin{align}
[{{\bm g}}_0]_{n_y,n_z}=\sqrt{a_{n_y,n_z}}e^{\frac{-j 2 \pi }{\lambda} r_{n_y,n_z}}, \quad \forall n_y,~ \forall n_z .
\end{align}

On the other hand, \rev{due to the rich scattering environment at the side of user~1 (see Fig.~\ref{system}) as well as the long distance between user~1 and the AP/IRS, the AP$\rightarrow$user~1 channel ${h_1}$ and IRS$\rightarrow$user~1 channel ${{\bm g}}_1$ are assumed to follow the Rayleigh fading channel model}, i.e., 
\begin{align}\label{Rayleigh}
{h_1}\sim {\mathcal N_c }\left(0, \frac{\beta}{d_{h_1}^\alpha} \right), \quad {{\bm g}}_1\sim {\mathcal N_c } \left({\bm 0}, \frac{2\beta}{d_{{{\bm g}}_1}^\alpha} {\bm I}\right)
\end{align}
where $\beta$ stands for the reference path gain at the distance of 1 meter (m),
\rev{the factor of $2$ in the latter accounts for the half-space reflection of each IRS element,}
$\alpha$ accounts for the path loss exponent between the AP/IRS and user~1,
and $d_{h_1}$ and $d_{{{\bm g}}_1}$ denote the reference distances between the AP/IRS and user~1, respectively.

For any given beamforming vector ${\bar{\bm \theta}}$ designed for user~2 (to be specified in Section \ref{Beamforming}), ${\rm diag}\left({\bar{\bm \theta}}\right)$ is an independent diagonal unitary matrix that will not change the distribution of ${{\bm g}}_1$. As such, it follows that ${\rm diag}\left({\bar{\bm \theta}}\right){{\bm g}}_1$ has the same distribution as ${{\bm g}}_1$, i.e., ${\rm diag}\left({\bar{\bm \theta}}\right){{\bm g}}_1\sim {\mathcal N_c }({\bm 0}, \frac{2\beta}{d_{{{\bm g}}_1}^\alpha} {\bm I})$. Accordingly, it can be shown that ${\bar g}_1= {{\bm g}}_0^T  {\rm diag}\left({\bar{\bm \theta}}\right) {{\bm g}}_1$ follows the Rayleigh fading channel model, i.e., ${\bar g}_1\sim {\mathcal N_c }(0, \varrho^2_{{\bar g}_1})$
with the average channel gain given by
\begin{align}\label{average_CH}
\varrho^2_{{\bar g}_1}&=\sum_{n_y=-\frac{N_y-1}{2}}^{n_y=\frac{N_y-1}{2}} 
\sum_{n_z=-\frac{N_z-1}{2}}^{n_z=\frac{N_z-1}{2}} 
a_{n_y,n_z}
\frac{2\beta}{d_{{{\bm g}}_1}^\alpha}=\frac{A\beta}{2\pi r^2 d_{{{\bm g}}_1}^\alpha } \sum_{n_y=-\frac{N_y-1}{2}}^{n_y=\frac{N_y-1}{2}} 
\sum_{n_z=-\frac{N_z-1}{2}}^{n_z=\frac{N_z-1}{2}}  
\frac{1}{\left(1+ (n_y^2+n_z^2)\epsilon^2\right)^{3/2}}.
\end{align}
Furthermore, by denoting $\varrho^2_{{h_1}}\triangleq \frac{\beta}{d_{h_1}^\alpha}$ as the average channel gain of the direct link, we can readily obtain the average received SNR in \eqref{SNR} as
\begin{align}\label{aveSNR}
{\bar \gamma}={\mathbb E}\{\gamma\}={\mathbb E}\left\{\frac{P_1(|{h_1}|^2+|{\bar g}_1|^2)}{\sigma^2}\right\}
={\bar P}_1\left( \varrho^2_{{h_1}} + \varrho^2_{{\bar g}_1}\right)
\end{align}
where ${\bar P}_1= \frac{P_1}{\sigma^2}$ is the transmit SNR for user~1.
\subsubsection{Average Channel Gain}
We first derive a simple closed-form expression for the average channel gain of the IRS-reflected link and then analyze its asymptotic performance as the number of IRS elements goes to infinity.

\indent\emph{Theorem 1}: With the IRS-aided transmit diversity and under the practical condition of $\epsilon\ll 1$, i.e., $\Delta\ll r$, the resultant average channel gain in \eqref{average_CH} can be expressed in a
closed-form as
\begin{align}\label{gain}
\varrho^2_{{\bar g}_1}=\frac{2\xi\beta}{\pi d_{{{\bm g}}_1}^\alpha } \arctan\left(\frac{N_y N_z \epsilon^2}{2\sqrt{4+(N_y^2+N_z^2)\epsilon^2}}\right).
\end{align}
\begin{IEEEproof} 
	Please refer to Appendix \ref{AppendixA}.
\end{IEEEproof}

\indent\emph{Remark 1}: According to Theorem 1, it is observed that the average channel gain in \eqref{gain} monotonically increases with the number of IRS elements $N$ (i.e., $N_y$ and/or $N_z$), but not linearly with $N$ as in the conventional uniform-plane wave model when the IRS is deployed at the user side without any passive beamforming gain \cite{Wu2019TWC}.
 Moreover, given an element spacing $\Delta$ and substituting $\epsilon=\frac{\Delta}{r}$ into \eqref{gain}, it can be verified that the average channel gain in \eqref{gain} increases as the AP-IRS distance $r$ decreases. This is expected since the propagation loss in the IRS-reflected link can be reduced by decreasing the AP-IRS distance.
As such, it is preferable to have the co-located AP and IRS for achieving a higher average channel gain over the IRS-reflected link.

Next, based on Theorem 1, we obtain the following lemma for the asymptotic performance on the average channel gain of the IRS-reflected link.

\indent\emph{Lemma 1}: For the infinitely large IRS with ${\bar N}\rightarrow \infty$, the channel gain in \eqref{gain} reduces to 
\begin{align}\label{limit}
\lim_{{\bar N}\rightarrow \infty} \varrho^2_{{\bar g}_1}=\frac{2\xi\beta}{\pi d_{{{\bm g}}_1}^\alpha }\times \frac{\pi}{2}=\frac{\xi\beta}{d_{{{\bm g}}_1}^\alpha },
\end{align}
where ${\bar N}$ is the base number of reflecting elements for each dimension,
and we denote $N_y=\eta_y{\bar N}$ and $N_z=\eta_z{\bar N}$ with $\eta_y$ and $\eta_z$ being the ratios along the $y$- and $z$-axis, respectively.
\begin{IEEEproof}
	Let $x=\frac{N_y N_z \epsilon^2}{2\sqrt{4+(N_y^2+N_z^2)\epsilon^2}}$ and it can be readily verified that $x\rightarrow \infty$ as ${\bar N}\rightarrow \infty$. Moreover, the asymptotic value of the function $\arctan\left(x\right)$ is given by $\lim\limits_{x\rightarrow \infty}\arctan\left(x\right)=\frac{\pi}{2}$, thus completing the proof.
\end{IEEEproof}

\begin{figure}[!t]
	\centering
	\includegraphics[width=2.8in]{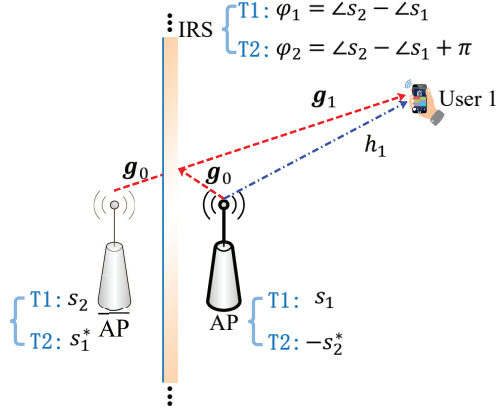}
	\setlength{\abovecaptionskip}{-3pt}
	\caption{Transmit diversity with a large-scale IRS and its equivalence to the Alamouti’s scheme.}
	\label{IRS_mirror}
	\vspace{-0.7cm}
\end{figure}

Lemma 1 implies that with an infinitely large IRS, the average channel gain of the IRS-reflected link approaches to a constant value of $\frac{\xi\beta}{d_{{{\bm g}}_1}^\alpha }$, rather than increasing unbounded.
This result makes intuitive sense since at most half of the AP transmit power can be reflected by the infinitely large IRS.
In particular, for the extreme case with the full array occupation ratio, i.e., $\xi=1$,  the IRS becomes like a mirror (see Fig.~\ref{IRS_mirror}) that reflects half of the AP transmit power in its front half-space reflection area and the asymptotic average channel gain of the IRS-reflected link is $\varrho^2_{{\bar g}_1}=\frac{\beta}{d_{{{\bm g}}_1}^\alpha }$.
Furthermore, since $d_{h_1} \approx d_{{{\bm g}}_1}$ for the distances from the co-located AP/IRS to user~1, we have  $\varrho^2_{{\bar g}_1}\approx \varrho^2_{{h_1}}$, which implies that the average channel gain of the IRS-reflected link is comparable to that of the direct link for achieving the desired transmit diversity performance. 
Intuitively, we can regard the sufficiently large IRS with $\xi=1$ as a sufficiently large mirror shown in Fig.~\ref{IRS_mirror}, where the $\overline{\text {AP}}$ is the mirror/image point of the AP with respect to the IRS. In this case, it appears that there are two ``active" antennas at the AP and $\overline{\text {AP}}$ to jointly achieve the transmit diversity similar to the Alamouti's scheme, where the mirror/image point (i.e., $\overline{\text {AP}}$) is able to transmit the sequence $\left[s_2, s_1^*\right]^T$ by \emph{equivalently} tuning the IRS common phase-shift $\varphi_i$ on a per-symbol basis (see Table \ref{coding}) .

\subsubsection{Average SER}
Next, we analyze the average SER of user~1 under the IRS-aided transmit diversity scheme.
Since both ${h_1}$ and ${\bar g}_1$ follow the Rayleigh distribution,
the received SNR  $\gamma$ in \eqref{SNR} becomes a central chi-square
distributed random variable with four degrees of freedom and the corresponding moment-generating function (MGF) is given by
\begin{align}\label{MGF}
{\mathbb M}_{\gamma} (x)=\left(1-x{\bar P}_1\varrho^2_{{h_1}}\right)^{-1}
\left(1-x{\bar P}_1\varrho^2_{{\bar g}_1}\right)^{-1} .
\end{align}
Based on \eqref{MGF}, we can calculate the average SER under the $M$-ary PSK constellation as \cite{goldsmith2005wireless}
\begin{align}\label{SER}
{\mathbb P}_e=\frac{1}{\pi}\int_{0}^{\frac{(M-1)\pi}{M}} {\mathbb M}_{\gamma} \left(-\frac{\sin^2 (\pi/M)}{\sin^2 (\phi)} \right) {\rm d}\phi.
\end{align}
Although the expression in \eqref{SER} does not lead to a closed-form solution, it can be easily evaluated by integrating over the finite range of $\left[0, \frac{(M-1)\pi}{M}\right]$ numerically, which is still more efficient than calculating the SER based on
the Monte-Carlo method.

To draw more useful insights, we are interested in determining a simple closed-form bound and the asymptotic performance for the average SER under the IRS-aided transmit diversity scheme. Specifically, by substituting \eqref{MGF} into \eqref{SER}, we have
\begin{align}\label{SER2}
{\mathbb P}_e=\frac{1}{\pi}\int_{0}^{\frac{(M-1)\pi}{M}} \left(\frac{\sin^2 (\phi)}{\sin^2 (\phi)+\sin^2 (\pi/M){\bar P}_1\varrho^2_{{h_1}} } \right) 
\left(\frac{\sin^2 (\phi)}{\sin^2 (\phi)+\sin^2 (\pi/M){\bar P}_1\varrho^2_{{\bar g}_1} } \right)
{\rm d}\phi.
\end{align}
Since $0\le\sin^2 (\phi)\le 1$, it is possible to obtain an upper bound on ${\mathbb P}_e$ by replacing $\sin^2 (\phi)$ with its minimum value, i.e., $0$ in the denominator of the integrand of \eqref{SER2}, yielding
\begin{align}\label{SER3}
{\mathbb P}_e\le  \frac{1}{{\bar P}_1^2 \pi\sin^4 (\pi/M)\varrho^2_{{h_1}}\varrho^2_{{\bar g}_1} }
\int_{0}^{\frac{(M-1)\pi}{M}} \sin^4 (\phi) {\rm d}\phi.
\end{align}
By applying the integral formula in \cite[eq. (2.513.7)]{gradshteyn2014table} to the integral part in \eqref{SER3}, we have
\begin{align}
C\triangleq \int_{0}^{\frac{(M-1)\pi}{M}} \sin^4 (\phi) {\rm d}\phi=\frac{3(M-1)\pi}{8M}-\frac{\sin\left(\frac{2(M-1)\pi}{M}\right)}{4}+\frac{\sin\left(\frac{4(M-1)\pi}{M}\right)}{32}
\end{align}
and thus
\begin{align}\label{bound}
{\mathbb P}_e\le  \frac{C}{ \pi\sin^4 (\pi/M)\varrho^2_{{h_1}}\varrho^2_{{\bar g}_1} }\cdot \frac{1}{{\bar P}_1^2}.
\end{align}

From \eqref{bound}, we observe that a transmit diversity of order two
is achieved. 
In particular, \rev{by dynamically adjusting the common phase-shift in the IRS-aided transmit diversity scheme, we not only refine the channel distribution (i.e., from the Rayleigh distribution to the chi-square distribution), but also
create the orthogonal channel condition for the information transmission of $s_1$ and $s_2$ to achieve the transmit diversity at the AP over the direct and reflected channels.}

\section{Passive Beamforming}\label{Beamforming}
While serving user~1 via the IRS-aided transmit diversity mode, the IRS-integrated AP can achieve the conventional passive beamforming function simultaneously for the other users with their CSI known.
In this section, we focus on the IRS passive beamforming design for user~2, where
the communication design and asymptotic performance are studied. 
\subsection{Passive Beamforming Design}\label{BFdesign}
Given the common IRS phase-shift to achieve transmit diversity for user~1, the beamforming vector ${\bar{\bm \theta}}$ in \eqref{decomposition} needs to be designed for user~2 to maximize the passive beamforming gain in the IRS-reflected (i.e., AP$\rightarrow$IRS$\rightarrow$user~2) channel, as shown in Fig.~\ref{system}.
Specifically, during each symbol period, we let ${\tilde s}$ denote the modulated data symbol for user~2 and
${\bm \theta}_i=e^{j\varphi_i}{\bar{\bm \theta}}$ denote the IRS reflection vector with $\varphi_i$ being the time-varying common phase-shift.
It is noted that the common phase-shift $\varphi_i$ is determined by the IRS-aided transmit diversity scheme for user~1 and thus is known at the IRS-integrated AP. As such, we let ${\tilde s}'_i=e^{-j\varphi_i}{\tilde s}$ be the pre-compensated modulated symbol to be transmitted by the AP, with $e^{-j\varphi_i}$ for offsetting the time-varying effect of the IRS common phase-shift.
Accordingly, the received signal at user~2 over another orthogonal frequency sub-band can be expressed as
\begin{align}\label{S1_LMU}
{\tilde y}&=\sqrt{P_2}\left({h_2} +{{\bm g}}_0^T  {\rm diag}\left({\bm \theta}_i\right) {{\bm g}}_2\right){\tilde s}'_i+{\tilde n}\notag\\
&=\sqrt{P_2}\Big({h_2}e^{-j\varphi_i} {\tilde s} +\underbrace{{{\bm g}}_0^T  {\rm diag}\left( {{\bm g}}_2 \right)}_{\bar{{\bm g}}_2^H} {\bar{\bm \theta}} e^{j\varphi_i} e^{-j\varphi_i} {\tilde s}\Big)+{\tilde n}\notag\\
&=\sqrt{P_2}\left({h_2}e^{-j\varphi_i}+{\bar g}_2 \left({\bar{\bm \theta}}\right) \right){\tilde s}+{\tilde n}
\end{align}
where ${P_2}$ denotes the AP transmit power for user~2,
 $\bar{{\bm g}}_2^H$ denotes the cascaded AP$\rightarrow$IRS$\rightarrow$user~2 channel (without taking the effect of IRS phase-shift yet),
  ${\bar g}_2 \left({\bar{\bm \theta}}\right) \triangleq \bar{{\bm g}}_2^H {\bar{\bm \theta}} $ denotes the effective complex-valued gain of the IRS-reflected channel,
  and ${\tilde n}$ is the zero-mean AWGN with variance $\sigma^2$ at user~2. 
  In particular, for any given beamforming vector ${\bar{\bm \theta}}$, we have
  \begin{align}
  \left|{\bar g}_2 \left({\bar{\bm \theta}}\right)\right|^2=\left|\bar{{\bm g}}_2^H {\bar{\bm \theta}}\right|^2=\left|\bar{{\bm g}}_2^H {\bar{\bm \theta}}e^{j\varphi_i}\right|^2=\left|\bar{{\bm g}}_2^H {\bm \theta}_i\right|^2.
  \end{align}
  This indicates that the IRS passive beamforming gain in any (channel) direction is invariant to the common phase-shift $\varphi_i$ applied to all of the reflecting elements.
  As such, to maximize the IRS passive beamforming gain in $\left|{\bar g}_2 \left({\bar{\bm \theta}}\right)\right|^2$ for user~2, the optimal beamforming vector can be designed as 
  \begin{align}\label{design}
  {\bar{\bm \theta}^{\star}}\triangleq\arg~\max_{{\bar{\bm \theta}}} \left|{\bar g}_2 \left({\bar{\bm \theta}}\right)\right|^2= e^{j\angle\left( \bar{{\bm g}}_2 \right)}.
  \end{align}
  By substituting \eqref{design} into \eqref{S1_LMU}, the received signal at user~2 can be further written as 
  \begin{align}\label{S1_LMU2}
  {\tilde y}=\sqrt{P_2} \left({h_2}e^{-j\varphi_i}+{\bar g}_2^{\star}\right) {\tilde s} +{\tilde n}
  \end{align}
  where we denote ${\bar g}_2^{\star} \triangleq {\bar g}_2 \left({\bar{\bm \theta}^{\star}}\right)=\left\| \bar{{\bm g}}_2\right\|_1$.
Accordingly, the received SNR at user~2 is given by 
\begin{align}\label{SINR}
{\tilde \gamma}=\frac{P_2|{h_2}e^{-j\varphi_i}+{\bar g}_2^{\star}|^2}{\sigma^2}.
\end{align}

\indent\emph{Remark 2}: From \eqref{SINR}, we notice that $e^{-j\varphi_i}$ will incur the phase-shift rotation to the direct channel ${h_2}$ and thus cause
the time-varying perturbation to the received SNR at user~2.
However, as will be shown later, \rev{by leveraging a large number of passive reflecting elements at the IRS close to the AP and performing the fine-grained passive beamforming toward user~2, the effective gain of the IRS-reflected channel will overwhelm that of the direct channel, i.e., $|{\bar g}_2^{\star}|^2\gg |{h_2}|^2=|{h_2}e^{-j\varphi_i}|^2$,} thus making this time-varying perturbation negligible under the passive beamforming mode.
On the other hand, \rev{the fine-grained IRS passive beamforming design requires accurate CSI or high-resolution beam training, which generally comes at the cost of relatively high training overhead.
Specifically, for the optimal passive beamforming design given in \eqref{design}, we need to acquire the full cascaded CSI of $\bar{{\bm g}}_2$ at the AP/IRS, which can be obtained via the existing
cascaded channel estimation schemes (see, e.g., \cite{wu2020intelligent,zheng2021survey,qingqing2019towards,zheng2019intelligent,zheng2020intelligent,yang2019intelligent,Zheng2020Fast}).
However, the training overhead is generally proportional to the number of reflecting elements $N$ and thus may incur a long estimation delay before data transmission.
As such, the passive beamforming mode is more favorable for boosting the communication rate of quasi-static/low-mobility users that are more tolerable to channel estimation delay with sufficiently long channel coherence interval.}


\subsection{Performance Analysis}\label{Analysis2}
In this subsection, we analyze the IRS passive beamforming gain for user~2. 
To be consistent, we consider the same UPA model for the IRS as shown in Fig.~\ref{Geometry}.
For ease of exposition, we assume that user~2 is located along the $x$-axis with its location denoted by ${\bm p}_{U2}=[{\tilde r},0,0]$ and ${\tilde r}>0$.\footnote{The results of this paper can be extended to the case where user~2 is arbitrarily located in the front half-space reflection area of the IRS, by taking into account the geometric angle formed by the location of user~2 and the IRS normal direction.}
\rev{Moreover, for simplicity, we assume the channel between the IRS and user~2 is LoS-dominant.
Similar to \eqref{AP_IRS}, the channel power gain between user~2 and the $(n_y,n_z)$-th IRS element is
given by}
\begin{align}\label{user_IRS}
b_{n_y,n_z}=\frac{A{\tilde r}}{4\pi {\tilde r}_{n_y,n_z}^3}=\frac{A}{4\pi {\tilde r}^2\left(1+ (n_y^2+n_z^2){\tilde \epsilon}^2\right)^{3/2}}
\end{align}
where ${\tilde r}_{n_y,n_z}\triangleq {\tilde r}\sqrt{1+ (n_y^2+n_z^2){\tilde \epsilon}^2}$ denotes the distance between user~2 and the center of the $(n_y,n_z)$-th IRS element, and ${\tilde \epsilon}\triangleq\frac{\Delta}{\tilde r}$ with ${\tilde \epsilon}\ll 1$.
Accordingly, the channel coefficient between user~2 and the $(n_y,n_z)$-th IRS element in ${{\bm g}}_2$ is given by
\begin{align}
[{{\bm g}}_2]_{n_y,n_z}=\sqrt{2b_{n_y,n_z}}e^{\frac{-j 2 \pi }{\lambda} {\tilde r}_{n_y,n_z}}, \quad\forall n_y,~ \forall n_z 
\end{align}
\rev{where the factor of $2$ accounts for the half-space reflection of each IRS element.}
Furthermore, with the optimal beamforming vector ${\bar{\bm \theta}^{\star}}$ given in \eqref{design}, the maximum passive beamforming gain can be expressed as
\begin{align}\label{passive_gain}
 \left|{\bar g}_2^{\star}\right|^2=
\left(\sum_{n_y=-\frac{N_y-1}{2}}^{n_y=\frac{N_y-1}{2}} 
\sum_{n_z=-\frac{N_z-1}{2}}^{n_z=\frac{N_z-1}{2}} 
\sqrt{2a_{n_y,n_z} b_{n_y,n_z}}\right)^2.
\end{align}
For notational convenience, we define the distance ratio as $\rho\triangleq \frac{r}{\tilde r}$ and obtain $\rho\ll 1$, i.e., $r\ll {\tilde r}$ due to the fact that user~2 is far away from the co-located AP/IRS in practice. Then we have the following theorem for the maximum passive beamforming gain.

\indent\emph{Theorem 2}: Under the practical condition of $\epsilon\ll 1$ and $\rho\ll 1$, the maximum passive beamforming gain in \eqref{passive_gain} is lower-/upper-bounded by
\begin{align}\label{passive_gain2}
\frac{2\rho}{1-\rho^2}\xi^2{\mathbb G}^2\left(R_L\right) \le \left|{\bar g}_2^{\star}\right|^2 \le 
\frac{2\rho}{1-\rho^2}\xi^2{\mathbb G}^2 \left(R_U\right)
\end{align}
where $R_L=\frac{1}{2}\epsilon \min \{N_y, N_z\}$, $R_U=\frac{1}{2}\epsilon\sqrt{N_y^2+ N_z^2}$, and
\begin{align}\label{GR}
{\mathbb G} \left(R\right)=\frac{1}{\sqrt{1+\sqrt{\frac{1}{\rho^2}-1} \cos \left(\arctan\left(R\right)\right)}}-
\frac{1}{\sqrt{1+\sqrt{\frac{1}{\rho^2}-1}}}.
\end{align}
\begin{IEEEproof} 
	Please refer to Appendix \ref{AppendixB}.
\end{IEEEproof}

Next, based on Theorem 2, we have the following lemma for the asymptotic passive beamforming gain of the IRS-reflected link.

\indent\emph{Lemma 2}: For an infinitely large IRS with ${\bar N}\rightarrow \infty$, the asymptotic passive beamforming gain in \eqref{passive_gain} becomes
\begin{align}\label{asymptotic_passive}
\lim_{{\bar N}\rightarrow \infty}\left|{\bar g}_2^{\star}\right|^2 
=\frac{2\rho}{1-\rho^2}\xi^2\left(1- \frac{1}{\sqrt{1+\sqrt{\frac{1}{\rho^2}-1}}}\right)^2,
\end{align}
where ${\bar N}$ is defined in Lemma 1.

\begin{IEEEproof}
	Please refer to Appendix \ref{AppendixC}. 
\end{IEEEproof}

Lemma 2 implies that with an infinitely large IRS, the maximum passive beamforming gain also approaches to a constant value (similar to the average channel gain in the IRS-aided transmit diversity scheme), rather than increasing unbounded with $N$.
Recall that for the IRS-integrated AP, we have $\rho\ll 1$ due to the fact that user~2 is far away from the co-located AP/IRS in practice. Accordingly, it follows that $\rho^2 \ll 1 $ and $\frac{1}{\rho^2} \gg 1$, and thus we can further approximate the asymptotic passive beamforming gain in \eqref{asymptotic_passive} as
\begin{align}\label{IRS_user}
\lim_{{\bar N}\rightarrow \infty}\left|{\bar g}_2^{\star}\right|^2 \approx
2\rho\xi^2=\frac{2r}{\tilde r}\xi^2.
\end{align}
From \eqref{IRS_user}, we can infer that $\lim\limits_{{\bar N}\rightarrow \infty}\left|{\bar g}_2^{\star}\right|^2 \ll 1$ due to the fact that $\rho\ll 1$ and $\xi\le 1$.
This result also makes intuitive sense with an infinitely large IRS (i.e., $N$) , since the effective passive beamforming gain cannot exceed $1$ due to the propagation loss. Otherwise, if the received power keeps increasing with the array dimension of IRS,
it may even exceed the transmit power, which is impossible. \rev{Note that this asymptotic result is in sharp contrast to the so-called ``squared power gain" of IRS passive beamforming that scales {\it quadratically} with $N$, i.e., in the order of $N^2$ \cite{Wu2019TWC}, which, however, is only valid under the far-field propagation model with not-so-large IRS and sufficiently large distance between IRS and AP/user.}

On the other hand, as a comparison, we assume that the direct link between the AP and user~2 is also LoS-dominant and its channel gain is given by 
\begin{align}\label{AP_user}
|{h_2}|^2=\frac{A'}{4\pi ({\tilde r}-r)^2} \approx \frac{A'}{4\pi {\tilde r}^2}
\end{align}
where $A'$ denotes the effective physical size of the antenna equipped at user~2 with $A'\ll 1$ in practice and the approximation follows that $r\ll {\tilde r}$ in practice.

\indent\emph{Remark 3}: From \eqref{AP_user}, it is observed that the channel power gain of the direct link decays inversely with the squared distance ${\tilde r}^2$, which is in agreement with the free-space path loss model.
In contrast, it is observed from \eqref{IRS_user} that with an infinitely large IRS, the passive beamforming gain decays inversely with the distance ${\tilde r}$ only, i.e., having a lower decaying order with respect to the propagation distance ${\tilde r}$. 


\section{Simulation Results}\label{Sim} 
In this section, we present simulation results to examine
the performance of the proposed scheme of simultaneous transmit diversity and passive beamforming at the IRS-integrated AP in the downlink communication. Under the 3D Cartesian coordinate system shown in Fig.~\ref{Geometry}, the IRS is placed on the $y$-$z$ plane with its center at the origin, the AP equipped with a single isotropic antenna is placed along the $x$-axis with $r=0.5$~m, and user~1 is randomly located in the front half-space reflection area of the IRS with the distance $d_{{{\bm g}}_1}=80$~m; while user~2 is placed along the $x$-axis with $\tilde r$ to be specified later depending on the scenarios.  
Without loss of generality, we consider the square UPA model for the IRS by setting the equal number of reflecting elements per dimension along the $y$- and $z$-axis, i.e., $N_y=N_z={\bar N}$ with $\eta_y=\eta_z=1$.
Unless otherwise stated, the wavelength is $\lambda=0.05$~m, the element spacing is set as $\Delta=\lambda/2=0.025$~m, and the physical size of each reflecting element is set as $A=\Delta^2$
with the full array occupation ratio of $\xi=1$.
Without loss of generality, we assume the AP adopts the equal transmit power for the two users over the two frequency sub-bands, i.e., $P_1=P_2=P$ in the simulations. The noise power at the two users is set equal as $\sigma^2=-85$~dBm.

\subsection{Transmit Diversity Performance}
First, we focus on the IRS-aided transmit diversity for user~1.
The reference path gain at the distance of $1$~m is set as $\beta=-30$~dB and the path loss exponent is set as $\alpha=3$ for both ${h_1}$ and ${{\bm g}}_1$ in \eqref{Rayleigh}. The modulation order of data symbols for user~1 is set as $M=8$, i.e., the $8$-ary PSK modulation.

\begin{figure}[!t]
	\centering
	\includegraphics[width=3.5in ]{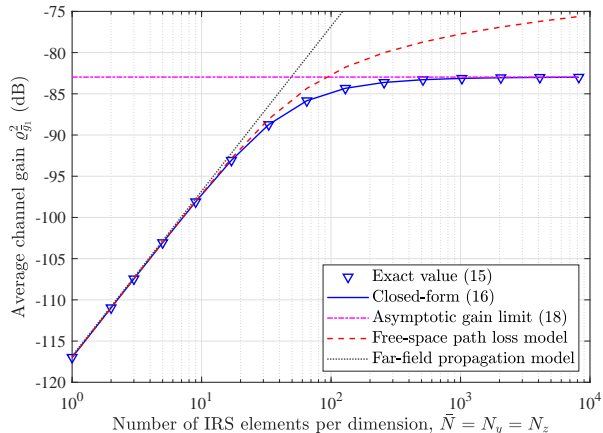}
	\setlength{\abovecaptionskip}{-5pt}
	\caption{Average channel gain versus the number of IRS elements per dimension, ${\bar N}=N_y=N_z$.}
	\label{Diversitygain_element}
	\vspace{-0.7cm}
\end{figure}
In Fig.~\ref{Diversitygain_element}, we plot the average channel gain  $\varrho^2_{{\bar g}_1}$ for the IRS-reflected link versus the number of IRS elements per dimension, ${\bar N}$. For comparison, we
consider the free-space path loss model (without accounting for the projected aperture) and the far-field propagation model (with the uniform-plane wave model) for the AP-IRS channel, by replacing \eqref{AP_IRS} with $a_{n_y,n_z}=\frac{A}{4\pi r_{n_y,n_z}^2}$ and $a_{n_y,n_z}=\frac{A}{4\pi r^2}$, respectively. The asymptotic gain limit of $\varrho^2_{{\bar g}_1}$ given in \eqref{limit} is also shown in the figure.  It is first observed that the closed-form expression derived in \eqref{gain} of Theorem~1 is in perfect agreement with the average channel gain $\varrho^2_{{\bar g}_1}$ calculated by \eqref{average_CH}. 
Besides, 
for a small to moderate number of IRS elements, the average channel gain $\varrho^2_{{\bar g}_1}$ increases
linearly with $N={\bar N}^2$ under all the channel models.
However, as ${\bar N}$ further increases, it is observed that due to the ignorance of the projected aperture and distance variation with the AP across different reflecting elements, both the free-space path loss model and far-field propagation model
tend to over-estimate the average channel gain. When ${\bar N}$ exceeds a certain threshold (say, ${\bar N}\ge 100$), the considered element-wise IRS channel model and the other two channel models exhibit drastically
different power scaling laws, i.e., approaching to a constant value versus increasing unbounded.

Next, for the SER performance comparison of the proposed IRS-aided transmit diversity scheme,
we consider the following three benchmark schemes.
\begin{itemize}
	\item {\bf Single-input single-output (SISO) transmission scheme without IRS}: In this scheme, a single-antenna AP sends modulated symbols directly to user~1.
	\item {\bf Dumb/Blind IRS-aided transmission scheme}: In this scheme, a single-antenna AP sends modulated symbols directly to user~1; while the IRS fixes its common phase-shift (which can be randomly generated following the uniform
	distribution within $[0, 2\pi)$ or designed to align the direct and reflected channels of user~2). 
	\item \rev{{\bf Classic Alamouti's scheme \cite{alamouti1998simple}}: In this scheme, a two-antenna AP sends the space-time code as in Table~\ref{coding} to user~1, with the average received SNR per symbol of ${\bar \chi}={\bar P}_1 \varrho^2_{{h_1}}$.}
	\item {\bf IRS-aided Alamouti's scheme \cite{khaleel2020reconfigurable}}: In this scheme, a single-antenna AP sends an unmodulated carrier signal and \rev{the IRS is partitioned into two equal-size subsurfaces to adjust their own common phase-shifts over any two symbol periods for emulating the space-time code in the classic Alamouti's scheme using the PSK modulation.} In this scheme, the unmodulated carrier signal through the direct channel is assumed to be perfectly canceled at user~1. As such, this scheme can also achieve a transmit diversity of order two and 
	 the average received SNR per symbol is given by ${\bar \zeta}={\bar P}_1 \varrho^2_{{\bar g}_1}$.
\end{itemize}

\begin{figure}[!t]
	\centering
	\includegraphics[width=3.5in ]{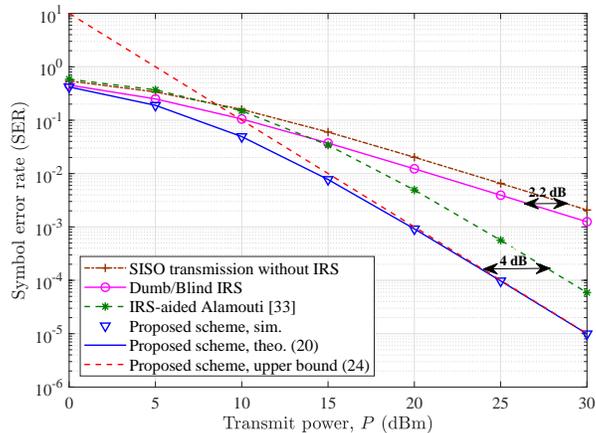}
	\setlength{\abovecaptionskip}{-5pt}
	\caption{\rev{SER versus transmit power $P$, with ${\bar N}=N_y=N_z=105$.}}
	\label{Alamouti_IRS_SNR}
	\vspace{-0.7cm}
\end{figure}

In Fig.~\ref{Alamouti_IRS_SNR}, we compare the average SER versus the transmit power for different schemes, with ${\bar N}=N_y=N_z=105$. Several interesting observations are made as follows. First, it is observed that for our proposed IRS-aided transmit diversity scheme, the analytical SER given in \eqref{SER} is in perfect agreement with the simulation result and the analytical upper bound given in \eqref{bound} is very tight when $P\ge 20$~dBm. Second, due to the additional signal reflection power induced by the IRS for achieving a higher average channel gain, the dumb/blind IRS-aided transmission scheme achieves up to $2.2$ dB gain over the conventional SISO transmission scheme without IRS. Third, as compared to the dumb/blind IRS-aided transmission scheme and the conventional SISO transmission scheme without IRS that have no transmit diversity, both the proposed scheme and the IRS-aided Alamouti's scheme in \cite{khaleel2020reconfigurable} \rev{achieve a transmit diversity gain of order two by dynamically adjusting the IRS common phase-shift and thus improve the
	SER performance significantly.} In addition, by jointly exploiting the direct and reflected channels for transmit diversity, \rev{the proposed scheme achieves up to $4$ dB gain over the IRS-aided Alamouti's scheme in \cite{khaleel2020reconfigurable} and $2.2$ dB gain over the classic Alamouti's scheme \cite{alamouti1998simple}, respectively, which also corroborates the accuracy of analytical results in Section~\ref{Analysis} as
$10\log_{10} \frac{{\bar \gamma}}{{\bar \zeta}}=10\log_{10}\left(1+\frac{\varrho^2_{{h_1}}}{\varrho^2_{{\bar g}_1}}\right)\approx 4~{\rm dB}$ and $10\log_{10} \frac{{\bar \gamma}}{{\bar \chi}}=10\log_{10}\left(1+\frac{\varrho^2_{{\bar g}_1}}{\varrho^2_{{h_1}}}\right)\approx 2.2~{\rm dB}$.}
Forth, at a low to medium power level (i.e., $P<15$ dBm), the IRS-aided Alamouti's scheme in \cite{khaleel2020reconfigurable} even performs worse than the dumb/blind IRS-aided transmission scheme. This can be explained by the fact that the direct link is ignored/canceled in IRS-aided Alamouti's scheme in \cite{khaleel2020reconfigurable}, thus suffering from a lower average channel gain.

\begin{figure}[!t]
	\centering
	\includegraphics[width=3.5in ]{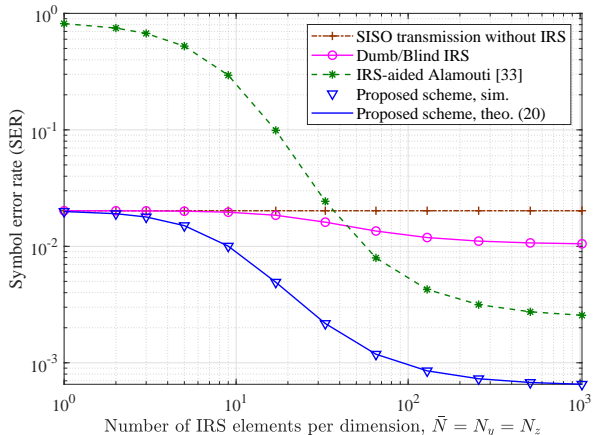}
	\setlength{\abovecaptionskip}{-5pt}
	\caption{\rev{SER versus the number of IRS elements per dimension ${\bar N}=N_y=N_z$, with $P=20$~dBm.}}
	\label{Alamouti_IRS_element}
	\vspace{-0.7cm}
\end{figure}
In Fig.~\ref{Alamouti_IRS_element}, we show the SER versus the number of IRS elements per dimension ${\bar N}$ for different schemes, with $P=20$~dBm. First, \rev{it is observed that the SERs of all the IRS-aided transmission schemes (except the classic Alamouti's scheme and the SISO transmission scheme without IRS) decrease as ${\bar N}$ increases,} which is attributed to the increased signal reflection power in the IRS-reflected link.
Second, the proposed scheme achieves the lowest SER among all the schemes. Third, when the number of IRS elements is not sufficiently large (i.e., ${\bar N}<50$), the IRS-aided Alamouti's scheme in \cite{khaleel2020reconfigurable} is even inferior to the dumb/blind IRS-aided transmission scheme and the conventional SISO transmission scheme without IRS;
while the IRS-aided Alamouti's scheme in \cite{khaleel2020reconfigurable} achieves a much lower SER
by increasing ${\bar N}$ and exploiting the transmit diversity gain. Nevertheless, due to the ignorance/cancellation of the direct link,
the IRS-aided Alamouti's scheme in \cite{khaleel2020reconfigurable} with ${\bar N}$ going to infinity still suffers from a constant performance gap as compared to the proposed IRS-aided transmit diversity scheme.

\subsection{Passive Beamforming Performance}

\begin{figure}[!t]
	\centering
	\includegraphics[width=3.5in ]{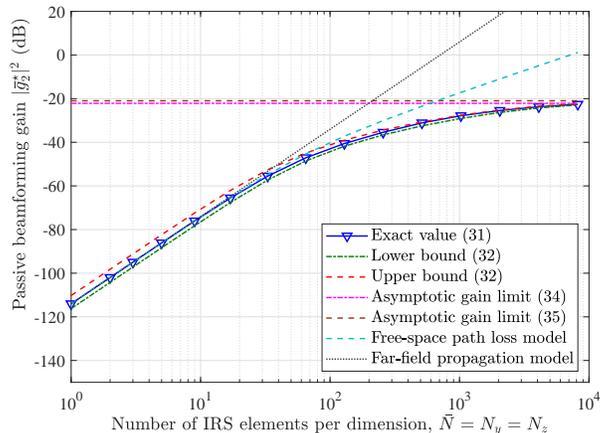}
	\setlength{\abovecaptionskip}{-5pt}
	\caption{Passive beamforming gain versus the number of IRS elements per dimension ${\bar N}$, with $P=20$~dBm and ${\tilde r}=50$~m.}
	\label{Beam_vs_IRSsize}
	\vspace{-0.7cm}
\end{figure}
Next, we focus on the IRS passive beamforming performance for user~2 by assuming that the transmit diversity for user 1 is being implemented at the same time.
In Fig.~\ref{Beam_vs_IRSsize}, we plot the maximum passive beamforming gain of the IRS-reflected link versus the number of IRS elements per dimension, ${\bar N}$.
Specifically, the exact value in \eqref{passive_gain}, the closed-form lower/upper bound in \eqref{passive_gain2}, and the asymptotic values in \eqref{asymptotic_passive} and \eqref{IRS_user} are shown in the figure.
First, one can observe that both the analytical lower and upper bounds in \eqref{passive_gain2} of Theorem 2 are quite tight and accurate to the exact passive beamforming gain calculated via \eqref{passive_gain}, especially with a large ${\bar N}$.
Moreover, as ${\bar N}$ further increases, the exact passive beamforming gain in \eqref{passive_gain}
exhibits a diminishing return and finally approaches to a constant value, which is also in accordance with the theoretical result in Lemma~2. On the other hand, \rev{we also consider the free-space path loss model (without accounting for projected aperture) with $a_{n_y,n_z}=\frac{A}{4\pi r_{n_y,n_z}^2}$ and $b_{n_y,n_z}=\frac{A}{4\pi {\tilde r}_{n_y,n_z}^2}$ and the far-field propagation model (with the uniform-plane wave model \cite{Wu2019TWC}) with $a_{n_y,n_z}=\frac{A}{4\pi r^2}$ and $b_{n_y,n_z}=\frac{A}{4\pi {\tilde r}^2}$ for comparison.
It is observed that with a small to moderate number of IRS elements per dimension (i.e., ${\bar N}\le 100$), the maximum passive beamforming gain
increases {\it quadratically} with $N$} and
achieves almost the same power gain under all the channel models. However, as ${\bar N}$ further increases,
\rev{both the free-space path loss model and the far-field propagation model \cite{Wu2019TWC} tend to {\it over-estimate} the actual performance by
the adopted element-wise IRS channel model in this paper, and their passive beamforming gains keep increasing unboundedly.}
In particular, for some large ${\bar N}$, the passive beamforming gains based on the free-space path loss model and the far-field propagation model even
exceed $1$, which is not possible.

\begin{figure}[!t]
	\centering
	\includegraphics[width=3.5in ]{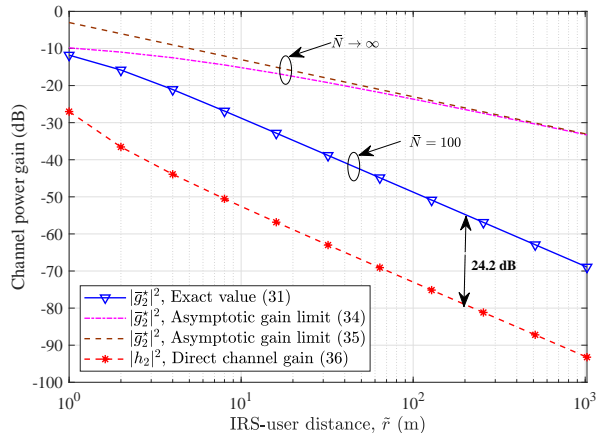}
	\setlength{\abovecaptionskip}{-5pt}
	\caption{Channel power gain versus the user distance ${\tilde r}$, with $P=20$~dBm.}
	\label{Beam_vs_distance}
	\vspace{-0.7cm}
\end{figure}
In Fig.~\ref{Beam_vs_distance}, we plot the channel power gain versus the user distance ${\tilde r}$ for the direct and reflected links.
Several interesting observations are made as follows. First, it is observed that as ${\bar N} \rightarrow \infty$,
the approximate asymptotic passive beamforming gain in \eqref{IRS_user} can be regarded as an upper bound for the asymptotic passive beamforming gain given in \eqref{asymptotic_passive}, which is very tight when ${\tilde r}\ge100$~m.
Second, all the channel power gains decrease as the user distance ${\tilde r}$ increases, which is expected since the overall channel power will decrease with the increased propagation distance ${\tilde r}$.
Third, as compared to the direct channel gain in \eqref{AP_user} that decays inversely with the squared distance ${\tilde r}^2$, both the asymptotic passive beamforming gains in \eqref{asymptotic_passive} and \eqref{IRS_user} have a lower decaying rate of order $1$ with respect to the distance ${\tilde r}$, which corroborates Remark~3 in Section~\ref{Analysis2}. 
However, with a finite number of IRS elements (e.g., ${\bar N}=100$ in Fig.~\ref{Beam_vs_distance}), the passive beam gain based on \eqref{passive_gain} still has a decaying rate of order $2$ with respect to the distance ${\tilde r}$, which is the same as that of the direct channel gain in \eqref{AP_user}. Despite the same order in decaying rate, the exact passive beam gain with a finite number of IRS elements ${\bar N}=100$ in \eqref{passive_gain} achieves an overwhelming power gain (up to $24.2$~dB) over the direct channel gain in \eqref{AP_user}, which also corroborates Remark~2 in Section~\ref{BFdesign}.

\begin{figure}[!t]
	\centering
	\includegraphics[width=3.5in ]{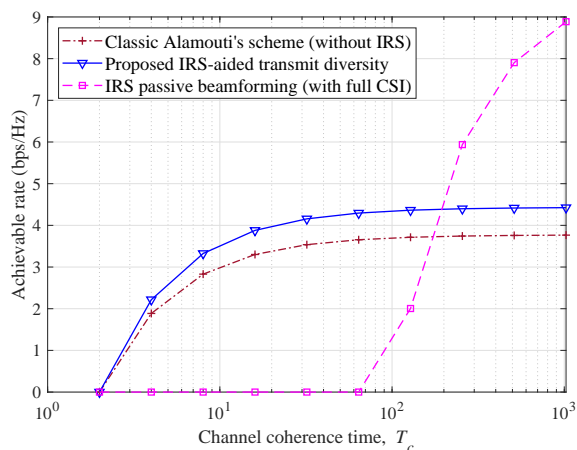}
	\setlength{\abovecaptionskip}{-5pt}
	\caption{\rev{Achievable rate versus channel coherence time, with $P=15$~dBm and ${\bar N}=N_y=N_z=100$.}}
	\label{Rate_vs_coherence}
	\vspace{-0.7cm}
\end{figure}
\rev{Finally, we plot the achievable rate versus the channel coherence time for different transmission schemes in Fig.~\ref{Rate_vs_coherence}. In particular, the effective achievable rate (with the training overhead taken into account) in bits per second per Hertz (bps/Hz) is defined as
\begin{align}
R\triangleq\left\{
\begin{aligned}
0,\qquad\qquad\qquad T_c\le T_0\\
\frac{T_c - T_0}{T_c}  {\mathbb E}\left\{\log_2\left(1+\gamma \right) \right\},\quad T_c> T_0
\end{aligned}
\right.\notag
\end{align}
where $\gamma$ represents the effective SNR with the practical estimation and modulation/coding taken into account,
$ T_c$ denotes the channel coherence interval, and $ T_0$ stands for the minimum training overhead.
For the IRS passive beamforming scheme, we group every $10 \times 10$ adjacent
IRS elements that share a common phase shift into a subsurface for design simplicity as in \cite{zheng2019intelligent}, which results in the minimum training overhead of $T_0=100+2$ symbol periods; while the minimum training overhead for the classic Alamouti's scheme and the proposed IRS-aided transmit diversity scheme is $2$ symbol periods. 
It is observed from Fig.~\ref{Rate_vs_coherence} that under the same AP transmit power and training overhead, the proposed IRS-aided transmit diversity scheme always achieves a higher effective rate than the classic Alamouti's scheme without IRS. This is expected since the AP transmit power can be reflected/recycled by the ``passive" IRS to improve the overall energy efficiency.
On the other hand, due to the high training overhead, the effective achievable rate of the IRS passive beamforming scheme is even lower than that of the other two schemes
when the channel coherence interval is relatively small (e.g., $T_c\le 200$ symbol periods).
This indicates that the IRS passive beamforming is more favorable for boosting the communication rate of quasi-static/low-mobility users with sufficiently long channel coherence time; while the IRS-aided transmit diversity is
highly appealing to the high-mobility/delay-sensitive communication with short channel coherence time.}

\section{Conclusions and Future Directions}\label{conlusion}
In this paper, we studied a new IRS-aided downlink communication system with the IRS-integrated AP, and proposed a new scheme of simultaneous transmit diversity and passive beamforming, by exploiting the fact that the IRS's passive beamforming gain in any (channel) direction is invariant to the common phase-shift applied to its reflecting elements for achieving transmit diversity.
We designed the common phase-shift of IRS elements to achieve transmit diversity at the AP side without the need of any CSI to serve high-mobility users; meanwhile, the IRS passive beamforming gain can be simultaneously achieved for serving low-mobility users with their CSI known at the AP.
Based on the adopted element-wise IRS channel modeling, we then analyzed the asymptotic performance of both IRS-aided transmit diversity and passive beamforming and derived the closed-form expressions to provide further insights as the number of IRS elements goes to infinity.
Numerical results validated our analysis and demonstrated the performance gains achieved by the proposed IRS-aided simultaneous transmit diversity and passive beamforming scheme, as compared to other benchmark schemes.

\rev{In the future, it is an interesting direction to study the more general cases of the multiple users, multi-antenna AP, and multi-IRS, which call
for more sophisticated system designs for the simultaneous transmit diversity and passive beamforming.
Moreover, for extending the above designs to the IRS-aided vehicular communications, how to jointly design adaptive passive beamforming and robust space-time coding for high-mobility users at the same time is also a non-trivial problem to solve. }

\appendices
\vspace{-0.5cm}
\section{Proof of Theorem 1}\label{AppendixA}
	 By exploiting the fact that $\epsilon \ll 1$ as in \cite{lu2021communicating,lu2021multi,feng2021wireless}, we approximate the double summation in \eqref{average_CH} with the double integral as
\begin{align}\label{average_CH2}
\varrho^2_{{\bar g}_1}=\frac{A\beta}{2\pi r^2 d_{{{\bm g}}_1}^\alpha } 
\frac{1}{\epsilon^2}
\int_{-\frac{N_y\epsilon}{2}}^{\frac{N_y\epsilon}{2}} 
\int_{-\frac{N_z\epsilon}{2}}^{\frac{N_z\epsilon}{2}} 
\frac{{\rm d}y {\rm d}z}{\left(1+ y^2+z^2\right)^{3/2}}
\end{align}
where we have 
\begin{align}
\frac{A\beta}{2\pi r^2 d_{{{\bm g}}_1}^\alpha } \frac{1}{\epsilon^2}= \frac{A\beta}{2\pi r^2 d_{{{\bm g}}_1}^\alpha } \frac{r^2}{\Delta^2}=\frac{\xi\beta}{2\pi d_{{{\bm g}}_1}^\alpha }.
\end{align}
By first integrating $y$ and then $z$,
the double integral in \eqref{average_CH2} can be calculated as
\begin{align}\label{Ddouble}
{\mathbb D}&\stackrel{(b1)}{=}\int_{-\frac{N_z\epsilon}{2}}^{\frac{N_z\epsilon}{2}} 
\frac{1}{1+z^2} 
\frac{y {\rm d}z}{\sqrt{1+ y^2+z^2}}\Bigg|_{-\frac{N_y\epsilon}{2}}^{\frac{N_y\epsilon}{2}} =\frac{1}{1+z^2} \frac{N_y\epsilon {\rm d}z}{\sqrt{1+ \frac{N_y^2\epsilon^2}{4}+z^2}}\notag\\
&\stackrel{(b2)}{=}2\arctan\left(\frac{N_y \epsilon z}{2\sqrt{1+ \frac{N_y^2\epsilon^2}{4}+z^2}}\right)\Bigg|_{-\frac{N_z\epsilon}{2}}^{\frac{N_z\epsilon}{2}}
=4\arctan\left(\frac{N_y N_z \epsilon^2}{2\sqrt{4+(N_y^2+N_z^2)\epsilon^2}}\right)
\end{align}
where $(b1)$ and $(b2)$ follow from the integral formulas in \cite[eq. (2.271.5)]{gradshteyn2014table} and \cite[eq. (2.284)]{gradshteyn2014table}, respectively.
By substituting \eqref{Ddouble} into \eqref{average_CH2}, we arrive at the expression given in \eqref{gain}, thus completing the proof.
\vspace{-0.5cm}
\section{Proof of Theorem 2}\label{AppendixB}
By substituting \eqref{AP_IRS} and \eqref{user_IRS} into \eqref{passive_gain} with ${\tilde \epsilon}^2=\rho^2\epsilon^2$, we have
\begin{align}\label{passive_gainSUM}
\left|{\bar g}_2^{\star}\right|
=\frac{\sqrt{2}A}{4\pi{\tilde r} r}
\sum_{n_y=-\frac{N_y-1}{2}}^{n_y=\frac{N_y-1}{2}} 
\sum_{n_z=-\frac{N_z-1}{2}}^{n_z=\frac{N_z-1}{2}} 
\frac{1}{ \left(1+ (n_y^2+n_z^2)\epsilon^2\right)^{3/4}}
\frac{1}{ \left(1+ (n_y^2+n_z^2)\rho^2\epsilon^2\right)^{3/4}}.
\end{align}
Similar to the proof of Theorem 1, we exploit the fact that $\epsilon \ll 1$ and approximate the double summation in \eqref{passive_gainSUM} with the double integral as
\begin{align}\label{passive_gainINT}
\left|{\bar g}_2^{\star}\right|
=\frac{\sqrt{2}A}{4\pi{\tilde r} r} \frac{1}{\epsilon^2}
\int_{-\frac{N_y\epsilon}{2}}^{\frac{N_y\epsilon}{2}} 
\int_{-\frac{N_z\epsilon}{2}}^{\frac{N_z\epsilon}{2}} 
\frac{{\rm d}y {\rm d}z}{ \left(1+ y^2+z^2\right)^{3/4} \left(1+\rho^2(y^2+z^2)\right)^{3/4}}
\end{align}
where we have 
\begin{align}
\frac{\sqrt{2}A}{4\pi{\tilde r} r} \frac{1}{\epsilon^2}=\frac{\sqrt{2}A}{4\pi{\tilde r} r} \frac{r^2}{\Delta^2}=\frac{\sqrt{2}\xi \rho}{4\pi}.
\end{align}
By applying a change of variables in the  polar coordinate system with ${\bar r}=\sqrt{y^2+z^2}$, we define the function ${\mathbb F} \left(R\right)$ as
\begin{align}\label{FR}
{\mathbb F} \left(R\right)\triangleq \int_{0}^{2 \pi} {\rm d}\omega
\int_{0}^{R} 
\frac{{\bar r} {\rm d}{\bar r}}{ \left(1+ {\bar r}^2\right)^{3/4} \left(1+\rho^2{\bar r}^2\right)^{3/4}}.
\end{align}
Notice that the double-integral area in \eqref{passive_gainINT} is rectangular, which is thus lower- and upper-bounded by its inscribed and circumscribed disks of radii $R_L=\frac{1}{2}\epsilon \min \{N_y, N_z\}$ and $R_U=\frac{1}{2}\epsilon\sqrt{N_y^2+ N_z^2}$, respectively. Accordingly, \eqref{passive_gainINT} is lower-/upper-bounded by
\begin{align}\label{bound2}
\frac{\sqrt{2}\xi \rho}{4\pi} {\mathbb F} \left(R_L\right)\le	\left|{\bar g}_2^{\star}\right| \le \frac{\sqrt{2}\xi \rho}{4\pi}  {\mathbb F} \left(R_U\right).
\end{align}
We let ${\bar r}=\tan \alpha $ and thus ${\mathbb F} \left(R\right)$ in \eqref{FR}  can be simplified as
\begin{align}\label{FR2}
{\mathbb F} \left(R\right)&=2 \pi
\int_{0}^{\arctan(R)} 
-\frac{  {\rm d}\cos \alpha}{ \left( \rho^2 +(1-\rho^2)\cos^2\alpha \right)^{3/4}} \notag\\
&\stackrel{(c1)}{\approxeq}2 \pi
\int_{0}^{\arctan(R)} 
-\frac{  {\rm d}\cos \alpha}{ \left( \rho^2 +2\rho\sqrt{1-\rho^2}\cos\alpha  +  (1-\rho^2)\cos^2\alpha \right)^{3/4}}\notag\\
&=2 \pi
\int_{0}^{\arctan(R)} 
-\frac{  {\rm d}\cos \alpha}{ \left( \rho  +\sqrt{1-\rho^2}\cos\alpha \right)^{3/2}}
\end{align}
\rev{where $(c1)$ is approximately obtained by exploiting the property of $\rho\ll 1$ such that the additive term $2\rho\sqrt{1-\rho^2}\cos\alpha$ in the denominator is negligible.} Moreover, we let $t=\cos\alpha$ and ${\mathbb F} \left(R\right)$ in \eqref{FR2} can be further simplified as
\begin{align}\label{FR3}
{\mathbb F} \left(R\right)&=2 \pi
\int_{\cos(\arctan(R))}^{1}
\frac{  {\rm d} t}{ \left( \rho  +\sqrt{1-\rho^2}t \right)^{3/2}}\stackrel{(c2)}{=}-\frac{4 \pi}{ \sqrt{1-\rho^2} \sqrt{\rho  +\sqrt{1-\rho^2}t} }\Bigg|_{\cos(\arctan(R))}^{1}\notag\\
&=\frac{4 \pi}{\sqrt{\rho}\sqrt{1-\rho^2}}
\underbrace{\left(\frac{1}{\sqrt{1+\sqrt{\frac{1}{\rho^2}-1} \cos \left(\arctan(R) \right)}}-
	\frac{1}{\sqrt{1+\sqrt{\frac{1}{\rho^2}-1}}}\right)}_{{\mathbb G}\left(R\right)}
\end{align}
where $(c2)$ is obtained by applying the integral formula in \cite[eq. (2.223.1)]{gradshteyn2014table}.
By substituting \eqref{FR3} into \eqref{bound2}, we can readily obtain \eqref{passive_gain2}, thus completing the proof. 
\vspace{-0.3cm}
\section{Proof of Lemma 2}\label{AppendixC}
As ${\bar N} \rightarrow \infty$, the radii of both  the
inscribed and the circumscribed disks (cf. Appendix \ref{AppendixB}) also go to infinity, i.e., $R_L, R_U\rightarrow \infty$. 
For ${\mathbb G} \left(R\right)$ given in \eqref{GR}, we have $\lim\limits_{R\rightarrow \infty}\arctan(R)=\frac{\pi}{2}$ and
$\lim\limits_{\alpha \rightarrow \frac{\pi}{2}}\cos \alpha =0$, hence
\begin{align}
\lim_{R\rightarrow \infty}{\mathbb G}\left(R\right)
=1- \frac{1}{\sqrt{1+\sqrt{\frac{1}{\rho^2}-1}}}.
\end{align}
According to the Squeeze Theorem,
it follows that both the lower- and upper-bounds of the passive beamforming gain in \eqref{passive_gain2}
approach to the same asymptotic value, i.e., 
\begin{align}\label{asymptotic}
\lim_{{\bar N}\rightarrow \infty}\left|{\bar g}_2^{\star}\right|^2 =
\lim_{R\rightarrow \infty}\frac{2\rho}{1-\rho^2}\xi^2{\mathbb G}^2\left(R\right)=\frac{2\rho}{1-\rho^2}\xi^2\left(1- \frac{1}{\sqrt{1+\sqrt{\frac{1}{\rho^2}-1}}}\right)^2
\end{align}
thus completing the proof.
\ifCLASSOPTIONcaptionsoff
  \newpage
\fi

\bibliographystyle{IEEEtran}
\bibliography{AlamoutiIRS}

\end{document}